\documentclass[aps,prl,showpacs,twocolumn,preprintnumbers,a4paper]{revtex4}
\usepackage[T1]{fontenc}
\usepackage[latin1]{inputenc}
\usepackage{graphicx}
\usepackage{amsmath}
\usepackage{amstext}
\usepackage{dsfont} 
\usepackage{textcomp} 

\renewcommand{\Im}{\mathrm{Im}}
\renewcommand{\Re}{\mathrm{Re}}
\usepackage{color} 

\makeatletter

\usepackage{graphicx}

\makeatother
\begin{document}
\preprint{CHO09, dated \today}

\title{Controlling synchrony by delay coupling in networks: from in-phase to splay and cluster states}

\author{Chol-Ung Choe$^{1,2}$}
\author{Thomas Dahms$^{1}$}
\author{Philipp H\"{o}vel$^{1}$}
\author{Eckehard Sch\"{o}ll$^{1}$}
\email{schoell@physik.tu-berlin.de}

\affiliation{
$^{1}$Institut f{\"u}r Theoretische Physik, Technische Universit{\"a}t Berlin,
10623 Berlin, Germany\\
$^{2}$Department of Physics, University of Science, Unjong-District, Pyongyang, DPR
Korea}

\begin{abstract}
We study synchronization in delay-coupled oscillator networks, using a master stability 
function approach. Within a generic model of
Stuart-Landau oscillators (normal form of super- or subcritical Hopf bifurcation) 
we derive analytical stability conditions and demonstrate that by tuning the coupling phase 
one can easily control the stability of synchronous periodic states. We propose the coupling phase
as a crucial control parameter to switch between 
in-phase synchronization or desynchronization for general network topologies, or between
in-phase, cluster, or splay states in unidirectional rings.
Our results are robust even for slightly nonidentical elements of the network.

\end{abstract}


\maketitle

Over the last decade, control of dynamical systems and stabilization of unstable states have become 
a central issue in nonlinear science \cite{SCH07}.
In parallel, the study of coupled systems ranging from a few elements to large networks 
has evolved into a rapidly expanding field \cite{WAT98}.
To determine the stability of synchronized oscillations in networks, Pecora and Carroll
introduced a technique called \textit{master stability function} (MSF) \cite{PEC98}, which allows one to
separate the local dynamics of the individual nodes from the network topology. 
Although some recent approaches have tried to extend this theory in the presence of time delays \cite{DHA04,KIN09}, 
up to now control and design of dynamic behavior in complex networks with time delay is still in its infancy. 

In this Letter, we aim to fill this gap by developing analytical 
tools for a large class of delay-coupled networks and deriving analytical 
conditions for controlling the different states of synchrony.
We identify the coupling phase as a crucial control parameter and demonstrate
that by adjusting this phase one can deliberately switch between different 
synchronous oscillatory states of the network.
We use a generic model describing a wide range of systems near a Hopf bifurcation,
which allows for an analytical treatment, including the calculation of 
the Floquet exponents.
These results promise broad applicability, since the presence of time delays is of 
crucial importance in a variety of physical, biological, technological, social, 
ecological, or economic networks where they occur, e.g., as 
propagation delays in communication networks and laser arrays \cite{FIS06,VIC08,FLU09}, 
electronic circuits \cite{RAM00},
neural systems \cite{GAS07b,BON07,SCH08}, or coupled Kuramoto phase oscillators 
\cite{EAR03,DHU08,SET08}, or in time-delayed feedback control loops \cite{PYR92}.

We consider $N$-dimensional networks of delay-coupled Stuart-Landau oscillators
($j=1,\ldots,N$) 
\begin{equation}
  \label{eq:network_adjacency}
 \dot{z}_j = f\left(z_j\right) + \sigma\sum_{n=1}^{N}a_{jn}\left(z_{n}(t-\tau)-z_j(t)\right),
\end{equation}
with $z_j = r_j e^{i\varphi_j} \in \mathds{C}$, time delay $\tau$, and complex coupling strength 
$\sigma=Ke^{i\beta}$. Such phase-dependent couplings have been shown to be important in overcoming
the odd-number limitation of time-delayed feedback control \cite{FIE07} and in anticipating 
chaos synchronization \cite{PYR08}.
The topology of the network is determined by the real-valued adjacency matrix  $\mathbf{A}=(a_{jn})$. 
Nonzero
diagonal elements, for instance, correspond to networks with delayed self-feedback. In the following, we consider only constant
row sum $\mu=\sum_n a_{jn}$ such that each node is subject to the same input for complete 
synchronization. This generalizes the common assumption of zero row sum in the MSF approach.
The local dynamics of each element is given by the normal form of a supercritical ($-$) or subcritical ($+$) Hopf bifurcation:
\begin{equation}
  \label{eq:local_hopf}
 f\left(z_j\right)=\big\lbrack{\lambda+i\omega\mp(1+i\gamma)\vert{z_j}\vert^2}\big\rbrack z_j
\end{equation}
with real constants $\lambda$, $\omega\neq 0$, and $\gamma$. This system
arises naturally as a generic expansion near a Hopf bifurcation, and is therefore 
often used as a paradigm for oscillators.

In the following, we focus on synchronous in-phase, cluster, and splay states with a common amplitude
$r_j \equiv r_{0,m}$ and phases given by 
$\varphi_j = \Omega_m t + j \Delta\phi_m$
with $\Delta \phi_m=2\pi m/N$.  
The integer $m$ determines the
specific state: in-phase oscillations correspond to $m=0$, while cluster and splay states correspond 
to  $m=1, \ldots, N-1$. The cluster number $d_c$, which determines how many clusters of oscillators exist, is given 
by the least common multiple of $m$ and $N$ divided by $m$.
$d_c=N$ corresponds to a splay state \cite{ZIL07}. 
With the above notation, we obtain
for in-phase oscillation in general networks and for splay and cluster states in ring configurations
\begin{subequations}\label{eq:amplitude_phase_2}
\begin{align}
r_{0,m}^2&=\pm\left(\lambda-\mu K\cos{\beta}+K\sum_{n=1}^{N}a_{jn}\cos{\Phi_{n,m}}\right)\\
\label{eq:Omega0}
\Omega_m &=\omega\mp \gamma r_{0,m}^2-\mu K\sin{\beta}+K\sum_{n=1}^{N}a_{jn}\sin{\Phi_{n,m}}
\end{align}
\end{subequations}
as invariant solutions of $r_{0,m}$ and $\varphi_j$ using the abbreviation $\Phi_{n,m}=\beta-\Omega_m\tau+(n-j)\Delta\phi_m$,
which is independent of $j$ in the cases mentioned above \cite{footnote}.
The following discussion focuses on the
supercritical case (upper signs), but a similar argument holds also for the subcritical Hopf normal form
(see discussion at the end of this Letter).

\begin{figure}[t]
  \includegraphics[width=\linewidth]{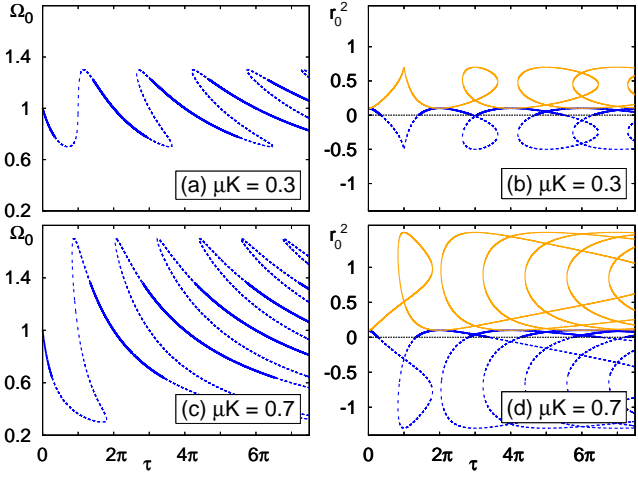}
  \caption{(Color online) Collective frequency $\Omega_0$ (left) and squared amplitude $r_0^2$ (right) 
of in-phase oscillation ($m=0$)
vs time delay $\tau$ for different amplitude of the feedback strength ($\mu K=0.3$ and $0.7$ 
in (a),(b) and (c),(d), respectively). Black (blue) and gray (yellow) curves correspond to a feedback phase $\beta=0$ and
$\beta=\Omega_0\tau$, respectively. Unphysical solutions ($r_0^2<0$) are dashed. 
For $\beta=\Omega_0\tau$ the curves in (a), (c) have the same shape, 
but no unphysical solutions occur. Parameters:
$\lambda=0.1$, $\omega=1$, $\gamma=0$.
}
  \label{fig:Omega_und_r0q}
\end{figure}

Figure~\ref{fig:Omega_und_r0q} shows solutions of $r_0^2$ ($r_{0} \equiv r_{0,0}$)  
and $\Omega_0$ for in-phase oscillations ($m=0$)
according to Eqs.~(\ref{eq:amplitude_phase_2}) in dependence on the time delay $\tau$ for fixed feedback strength
$\mu K=0.3$ and $0.7$ in panels (a),(b) and (c),(d), respectively. The black (blue) lines show the behavior
for the coupling phase $\beta=0$. The
collective frequency $\Omega_0$ is distributed around the intrinsic frequency $\omega=1$, where multiple solutions are obtained
with increasing time delay $\tau$. This behavior becomes more pronounced for higher $\mu K$ (c). The collective
amplitude also shows multivalued behavior; spurious solutions with $r_0^2<0$, which correspond to amplitude
death, are indicated as dashed curves.
For a coupling phase $\beta=\Omega_0\tau$, these unphysical solutions do not occur since
$r_0^2\geq\lambda=0.1$ as shown by the gray (yellow) curves in (b),(d). Note that for
$\beta=\Omega_0\tau$ the shape of the $\Omega_0$ curve in (a),(c) is 
unchanged, but now {\em all} points are valid solutions.

Considering small deviations $\delta r_j$ and $\delta\varphi_j$, i.e, $r_j=r_{0,m}(1+\delta r_j)$, $\varphi_j=\Omega_m
t+j \Delta\phi_m+\delta\varphi_j$, $\xi_j=\left(\delta r_j,\delta\varphi_j\right)^T$,
yields a variational equation for the
synchronized state
\begin{equation}
\mathbf{\dot{\xi}}=\mathbf{I}_N\otimes (\mathbf{J}_{0,m}^\mp-K\mathbf{\Psi}_m)\xi
+K(\mathbf{A}\otimes \mathbf{R}_{n,m})\mathbf{\xi}(t-\tau)
\label{eq:variational_matrixform}
\end{equation}
with the $2N$-dimensional vector $\xi=\left(\xi_1,\dots,\xi_N\right)^T$, 
the $N\times N$ identity matrix $\mathbf{I}_N$,
and matrices 
$\mathbf{\Psi}_m=\left(\begin{array}{cc}
\sum_na_{jn}\cos{\Phi_{n,m}}&-\sum_na_{jn}\sin{\Phi_{n,m}}\\
\sum_na_{jn}\sin{\Phi_{n,m}}&\sum_na_{jn}\cos{\Phi_{n,m}}
\end{array}\right)$
, 
$\mathbf{R}_{n,m}=\left(\begin{array}{cc}
\cos{\Phi_{n,m}}&-\sin{\Phi_{n,m}}\\
\sin{\Phi_{n,m}}&\cos{\Phi_{n,m}}
\end{array}\right)$
,
$\mathbf{J}_{0,m}^\mp=\left(\begin{array}{cc}
\mp2r_{0,m}^2&0\\
\mp2\gamma r_{0,m}^2&0
\end{array}
\right)$, 
which is an important generalization of the usual MSF approach. 

In order to derive an analytical expression for stability, Eq.~\eqref{eq:variational_matrixform} has to be diagonalized
in terms of $\mathbf{A}$. To succeed, the rotational matrix $\mathbf{R}_{n,m}$ must not depend on $n$. This is achieved
in two cases: (i) By considering only in-phase synchronization ($m=0$), or (ii) by considering special network
configurations. 

In case (i), the matrix $\mathbf{R}_{n,0}=\mathbf{R}$ with $\Phi_{n,0}=\Phi_0$
does not depend on $n$ and with $J_{0,0}^\mp \equiv J_0^\mp$ Eq.~\eqref{eq:variational_matrixform} simplifies to
\begin{equation}
\mathbf{\dot{\xi}}=\mathbf{I}_N\otimes (\mathbf{J}_{0}^\mp-\mu K\mathbf{R})\xi
+K(\mathbf{A}\otimes \mathbf{R})\mathbf{\xi}(t-\tau)
\end{equation}
Diagonalizing $\mathbf{A}$, we arrive at the block-diagonalized variational equation:
\begin{equation}
\mathbf{\dot{\zeta}}_k(t)=\mathbf{J}_{0}^\mp\mathbf{\zeta}_{k}(t)- K\mathbf{R}\left[\mathbf{\mu\zeta}_{k}(t)-\nu_k
\mathbf{\zeta}_k(t-\tau)\right],
\label{eq:fall_i_diagonalized}
\end{equation}
where $\nu_k$ is an eigenvalue of $\mathbf{A}$, $k=0,1, 2,\dots, N-1$, and $\nu_0=\mu$ corresponds to the dynamics in the synchronization manifold.
Since the coefficient matrices in Eq.~\eqref{eq:fall_i_diagonalized} do not depend on time, the Floquet 
exponents of the synchronized periodic state are given by the eigenvalues $\Lambda$ of the characteristic equation
\begin{equation}
\det\big{\lbrace}{\mathbf{J}_0^\mp-\Lambda\mathbf{I_2}+K\big(-\mu+\nu_k
e^{-\Lambda\tau}\big)\mathbf{R}}\big{\rbrace}=0.
\label{eq:fall_i_det}
\end{equation}

\begin{figure}[t]
   \includegraphics[width=\linewidth]{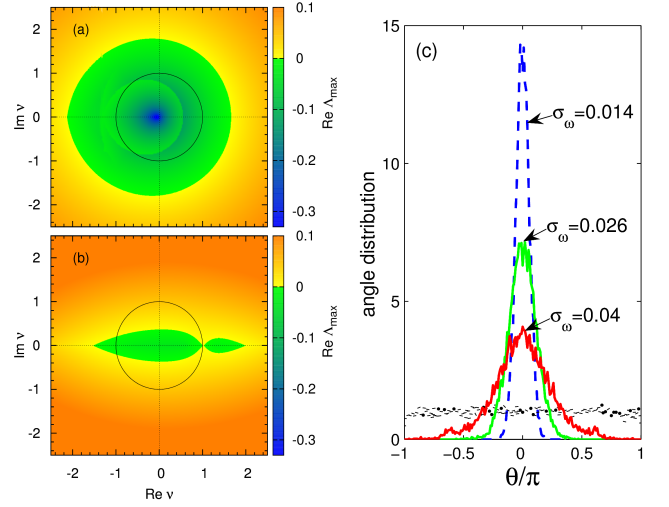}
  \caption{(Color online) (a), (b) Master stability function in the $(\Re\,\nu,\Im\,\nu)$-plane for
$m=0$, $\beta=0$. The gray scale (color code)
corresponds to the largest real part of the Floquet exponents for a given value of the product $K \nu$. All eigenvalues
of the coupling matrix $\mathbf{A}$ for a unidirectional ring lie on the black circle. 
Parameters:(a) $K\mu=0.3$, $\tau=2\pi$, (b) $K\mu=0.08$, $\tau=0.52\pi$, others as in Fig.\ref{fig:Omega_und_r0q}.
(c) Distribution of relative phases $\theta=\varphi_i-\Theta$ around the order parameter 
for $200$ slightly nonidentidal elements with different standard deviations $\sigma_{\omega}$ of the 
frequencies $\omega$ 
for $\beta=\Omega_0\tau$.  Dotted (black) curve: $\beta=\Omega_0\tau+\pi$ (desynchronization). 
Other parameters as in (b).
}
  \label{fig:msf}
\end{figure}

Figures~\ref{fig:msf}(a),(b) depict the MSF, i.e., the largest real part of the Floquet exponents, 
calculated from Eq.~\eqref{eq:fall_i_det} for different coupling parameters.
Note that for a unidirectionally coupled ring all eigenvalues, i.e., $\nu_k=\exp(2\pi i k/N)$ with
$k=0, 1, \dots, N-1$, are located on the black circle. Hence, for the choice of parameters in panel (a)
all eigenvalues lie in the region of negative maximum real part of the Floquet exponent 
(stable in-phase solution), 
whereas the parameters in panel (b) do not allow for synchronization of the unidirectional ring. 
Furthermore, it can be shown using Gerschg{\"o}rin's disk theorem \cite{EAR03}
that the eigenvalues are located on or inside this circle $S(0,\mu)$ centered at 
$0$ with radius $\mu$ for any network topology without self-feedback (diagonal elements $a_{jj}=0$).
The same holds for the circle $S(a_{jj},\mu)$ centered at $a_{jj}$ if self-feedback 
(constant $a_{jj}$, $j=1\dots, N$) is added while keeping the constant row sum condition. 
Note that the MSF is symmetric with respect to a change of sign of $\Im(\nu)$. 

For coupling phases $\beta=\Omega_0\tau+2l\pi$ with integer $l$, i.e., $\Phi_0= 2l\pi$, the characteristic
equation ~\eqref{eq:fall_i_det} for the Floquet exponents $\Lambda$ factorizes: 
\begin{equation}\label{eq:dominant} 
 0=\left(K\nu_k e^{-\Lambda\tau}-K\mu\mp2r_0^2-\Lambda\right)\left(K \nu_k
e^{-\Lambda\tau}-K\mu-\Lambda \right). 
\end{equation} 
In the supercritical case (upper sign), the dominant Floquet exponent is determined by the second factor
in Eq.~\eqref{eq:dominant} which gives Re $\Lambda<0$ for $K>0$ for any eigenvalue $\nu_k$
on or inside $S(0,\mu)$, taking into account $r_0^2>0$,
and hence stable in-phase synchronization for any network topology 
without or with self-feedback, observing the constant row sum conditions. 
A similar equation arises for
$\beta=\Omega_0\tau+(2l+1)\pi$ with $\Omega_0$ obtained from Eq.~(\ref{eq:Omega0}), 
and it can be shown
by analogous arguments that there exist exponents $\Lambda$ with positive real part, which results
in desynchronization. 
In conclusion, the synchronous (in-phase) dynamics can be stabilized or destabilized by proper choice 
of the coupling phase $\beta$. These results are robust even if slightly nonindentical elements are
considered. Fig.~\ref{fig:msf}(c) shows numerical simulations for 
all-to-all coupling of 200 elements with Gaussian frequency distributions around $\omega=1$ with 
different standard deviations $\sigma_\omega$. 
For $\beta=\Omega_0\tau$ the relative phases of the individual oscillators
around the order parameter 
$Re^{i\Theta}\equiv\frac{1}{N}\sum_{j}\frac{z_j}{\left|z_j\right|}$
(in rotating coordinates) are distributed around a maximum (in-phase)
with different sharpness according to $\sigma_{\omega}$,  
whereas for $\beta=\Omega_0\tau+\pi$ the phase is uniformly distributed 
regardless of $\sigma_{\omega}$ (desynchronization).

We now consider case (ii), i.e., special network configurations, and exemplarily choose a unidirectional
ring with $a_{i,i+1}=a_{N,1}=1=\mu$ and all other $a_{i,j}=0$. Then the matrix $\mathbf{R}_{n,m}=\mathbf{R}_{m}$ with
$\Phi_{n,m} = \Phi_m=\beta-\Omega_m\tau+\Delta\phi_m$ does not depend on $n$ and the diagonalization of
Eq.~\eqref{eq:variational_matrixform} yields the same form as Eq.~(\ref{eq:fall_i_diagonalized}) with
$\mu=1$, and $\mathbf{R}$ replaced by $\mathbf{R}_{m}$. 
The eigenvalues of $\mathbf{A}$ are explicitly given by $\nu_k=e^{2ik\pi/N}$, $k=0,1,\dots,N-1$, and
the Floquet exponents $\Lambda$ can be calculated from the corresponding characteristic equation
(\ref{eq:fall_i_det}).

\begin{figure}[t]
  \includegraphics[width=\linewidth]{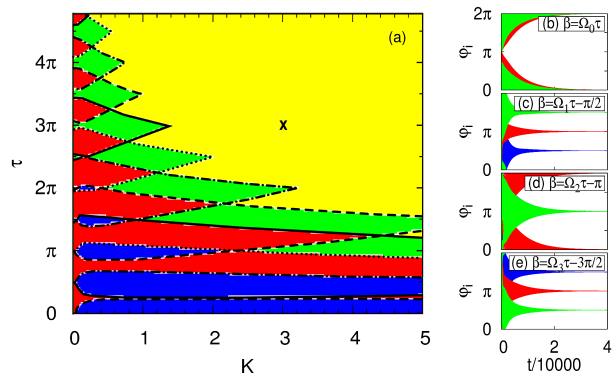}
  \caption{(Color online) (a) Stability diagram of unidirectional ring of $N=4$ oscillators
in the $(K,\tau)$-plane ($\beta=0$). 
Solid, dash-dotted, dashed, and dotted boundaries correspond to a stability change of 
in-phase ($m=0$), 2-cluster ($m=2$), splay states with $m=1$, and $m=3$, respectively. 
The black (blue), dark gray (red), light gray (green), and white (yellow) regions denote
multistability of one, two, three, and four of the above states.
(b)-(e) Time series of the phase differences for a unidirectional ring of four slightly nonidentical 
oscillators: 
(b) $\beta=\Omega_0\tau$, (c) $\Omega_1\tau-\pi/2$, (d) $\Omega_2\tau-\pi$, (e) $\Omega_3\tau-3\pi/2$,
with $\Omega_0=1$, $\Omega_1=0.83903$, $\Omega_2=1$, and $\Omega_3=1.16097$. Black (blue), dark
gray (red), and light gray (green) lines denote the differences 
$\varphi_2-\varphi_1$, $\varphi_3-\varphi_1$, and $\varphi_4-\varphi_1$, respectively 
(in (b),(d) black (blue) is hidden behind light gray (green)).  
Parameters: $\lambda=0.1$, $\gamma=0$, $K=3$, $\tau=3\pi$,
$\omega_1 = 0.99757$, $\omega_2 =0.99098$, $\omega_3 = 1.01518$ and $\omega_4 = 0.99496$.
}
  \label{fig:K_tau_ring4}
\end{figure}

Figure~\ref{fig:K_tau_ring4}(a) shows the stability boundaries of different dynamical scenarios
in the $(K,\tau)$-plane for unidirectional coupling of $N=4$ oscillators. The coupling phase is fixed at $\beta=0$. 
The gray scale (color code) indicates regions of different multistability of
in-phase ($m=0$), 2-cluster ($m=2$), and splay states ($m=1$, $m=3$):
Black (blue), dark gray (red), light gray (green), and yellow (white) color corresponds to regions 
where one, two, three, or four of these dynamical states are stable, respectively.

Let us now consider the effects of the coupling phase $\beta$. The specific choice of $\beta=\Omega_0\tau$,
$\Omega_1\tau-\pi/2$, $\Omega_2\tau-\pi$, and $\Omega_3\tau-3\pi/2$ enlarges the stability regime 
of the in-phase, splay ($m=1$),
cluster, and splay ($m=3$) states, respectively, to the complete $(K,\tau)$-plane. This can be understood as follows.
For $\Phi_m=2l\pi$, i.e., $\beta=\Omega_m\tau-\Delta\phi_m+2l\pi$ with integer 
$l$, the characteristic equation can again be factorized as Eq.(\ref{eq:dominant}).
For the supercritical case, taking into account $r_{0}^2>0$ at $\beta=\Omega_m\tau-\Delta\phi_m+2l\pi$,
it follows again that the dominant Floquet exponents 
have negative real part for any $K>0$ and $\tau$. Therefore the unidirectional ring configuration of 
Stuart-Landau
oscillators exhibits in-phase synchrony, splay state and clustering according to the choice of the 
control parameter
$\beta=\Omega_0\tau$, $\Omega_1\tau-2\pi/N$, or $\Omega_m\tau-2\pi m/N$ $(m>1, N>2)$, respectively, for any
values of the coupling strength and time-delay. 

To illustrate this further and demonstrate the robustness of our stability results for slightly
nonidentical oscillators, we choose a 
set of control parameters $K=3$ and $\tau=3\pi$, denoted by the black
cross in Fig.~\ref{fig:K_tau_ring4}(a), for which multistability of all four possible synchronization 
states is found for the coupling phase $\beta=0$.
Figures~\ref{fig:K_tau_ring4}(b)-(e) show time series from numerical simulations of four 
Stuart-Landau oscillators in a unidirectional
ring configuration with slightly different frequencies $\omega$.
For each choice of $\beta$ in panel (b) - (e) the solutions $\Omega_m$ were obtained by solving
Eqs.~\eqref{eq:amplitude_phase_2} such that the solution of $\Omega_m$ closest to unity was chosen. 
The differences of the phases $\varphi_i$ ($i=2,3,4$) relative to the first oscillator phase
$\varphi_1$ are plotted. After transients (note that the transient oscillations are not resolved on the 
time scale chosen), the oscillators behave exactly as predicted by our theory, 
i.e., they lock into in-phase synchronization for $\beta=\omega\tau$ (b), into
a splay state for $\beta=\omega\tau-\pi/2$ (c), into a 2-cluster state for
$\beta=\omega\tau-\pi$, where $\varphi_1=\varphi_3$ and $\varphi_2=\varphi_4$ (d), 
and again into a splay state, albeit with inverted ordering of the phases, 
for $\beta=\omega\tau-3\pi/2$ (e). 

Finally, for the subcritical Hopf normal form, 
it can be shown that the periodic orbit, which is unstable in the uncoupled case, can be 
stabilized in-phase synchronously by, e.g., bidirectional ring, star, or all-to-all coupling 
without self-feedback. In
these cases, both the synchronization manifold and the transversal modes are stable. 
There, Floquet exponents $\Lambda$
satisfy again Eq.~(\ref{eq:fall_i_det}) with $\nu_0=\mu$. For proper coupling phases $\beta$,
we find a finite interval of feedback gain $K$ for which the real parts of all Floquet 
exponents are negative. Note that
the in-phase synchronization manifold coincides with a single oscillator with delayed self-feedback 
as considered in Ref.~\cite{FIE07} to refute the alleged odd number limitation.

\begin{figure}[t]
  \includegraphics[width=\linewidth]{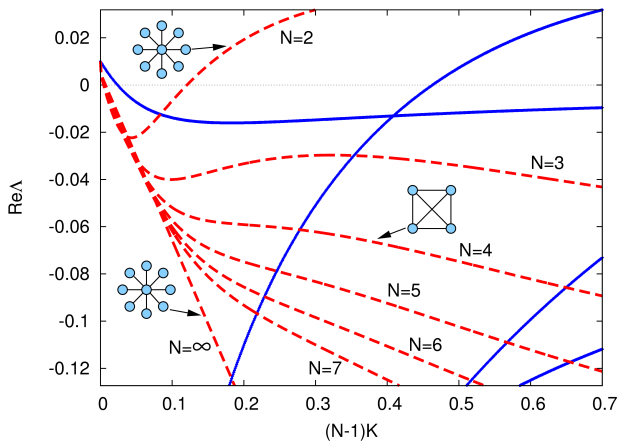}
\caption{(Color online) Real part of the Floquet exponents $\Lambda$ in the subcritical case vs reduced coupling
strength $(N-1)K$ in the all-to-all and star coupling configurations for different
$N$. The solid (blue) lines denotes Re$\Lambda$ inside the synchronization manifold for both all-to-all and star coupling. The dashed (red) lines show the largest 
transversal Re$\Lambda$ for different $N$ and all-to-all 
coupling. Floquet exponents for star coupling are independent of $N$. Parameters (as in \cite{FIE07}): $\lambda=-0.005$, $\omega= 1$, $\gamma=-10$, 
$\tau=2\pi/(1-\gamma\lambda)$, $\beta=\pi/4$.
}
\label{fig:subcritical}
\end{figure}

For all-to-all coupling the adjacency matrix is given by $a_{i,i}=0$ and $a_{i,j}=1$ for $i\neq j$ 
and $i,j=1,\ldots,N$, while for star coupling all $a_{i,j}=0$, except $a_{1,i}=1$ and 
$a_{i,1}=N-1$ for $i=2,\ldots,N$. 
Figure~\ref{fig:subcritical} shows Re $\Lambda$ in the subcritical case as a 
function of coupling strength $\mu K$ for all-to-all and star coupling. The solid (blue) lines 
corresponds to Re $\Lambda$ inside the synchronization manifold 
($\nu_0=\mu=N-1$ for both coupling configurations). 
For all-to-all coupling the transversal eigenvalues of $\mathbf{A}$ are given by 
$\nu_k=-1$ for $k=1,\ldots,N-1$ and the corresponding largest Re $\Lambda$ are 
denoted by the dashed (red) lines in Fig.~\ref{fig:subcritical} for different $N$. 
For star coupling, the transversal eigenvalues of $\mathbf{A}$ are $\nu_k=0$ 
for $k=1,\ldots,N-2$ and $\nu_{N-1}=-(N-1)$, and the corresponding largest Re $\Lambda$
are marked schematically.
We stress that for both all-to-all and star coupling there exists an interval of feedback strength
$K$ in which all Re $\Lambda<0$. Thus, time delayed coupling results in 
stabilization and in-phase synchronization.

In conclusion, we have shown that by tuning the coupling phase in delay-coupled networks 
one can easily control the stability
of synchronous periodic states, and we have specified analytic conditions. 
In general networks in-phase synchronization or desynchronization 
can be chosen, and in unidirectional rings either in-phase, cluster or splay states 
can be selected. The coupling phase is a parameter which is readily accessible,
e.g., in optical experiments \cite{SCH07}. Our results are robust even for slightly nonidentical elements of the network.

\begin{acknowledgments}
C.-U. C. acknowledges support from Alexander von Humboldt Foundation. This work was also supported by 
DFG in the framework of Sfb 555. We thank A. Amann for valuable discussions.
\end{acknowledgments}


\begin{thebibliography}{10}
\bibitem{footnote}
Note that $d_c$-clusters can also exist in more general networks, e. g., 
with circulant adjacency matrix.
\end{thebibliography}

\end{document}